\documentclass[twocolumn,showpacs,aps]{revtex4}
\usepackage{graphicx}
\begin{document}
\title{Spin-polaron model: transport properties of EuB$_6$}
\author{Jayita Chatterjee, Unjong Yu, and B. I. Min}
\affiliation{Department of Physics,
        Pohang University of Science and Technology,
        Pohang 790-784, Korea}
\date{\today}

\begin{abstract}

To understand anomalous transport properties of EuB$_6$,
we have studied the spin-polaron Hamiltonian incorporating the
electron-magnon and electron-phonon interactions.
Assuming a strong exchange interaction between carriers
and the localized spins,
the electrical conductivity is calculated.
The temperature and magnetic field
dependences of the resistivity of EuB$_6$ are well explained.
At low temperature, magnons dominate the
conduction process, whereas the lattice contribution becomes significant
at very high temperature due to the scattering with the phonons. Large
negative magnetoresistance near the ferromagnetic transition is also
reproduced as observed in EuB$_6$.
\end{abstract}

\maketitle

Hexaboride compounds have been studied extensively for their
unusual transport and magnetic properties. Among those,
EuB$_6$ has attracted special interest due to 
its exotic magnetic properties, such as two consecutive
phase transitions at low temperature (15.5 K and 12.6 K), 
very low carrier density ($\sim 10^{20}$ cm$^{-3}$), metallic ferromagnetism
below $T_c$, and large negative colossal magnetoresistance at the transition
of 15 K \cite{Fujita,Sullow}. 
It was proposed that the higher transition
temperature is a metallization temperature due to an increase in the
number of itinerant electrons and the lower one is a bulk Curie temperature
of long-range ferromagnetic (FM) order \cite{Sullow}. 
Sample dependence is an important issue
in EuB$_6$ as it has a very small number of intrinsic carriers. 
The higher transition temperature is seen to be much more 
sample dependent than the lower one \cite{Sullow}.  
 
Raman scattering measurements \cite{Nyhus,Snow} indicated the 
spontaneous formation of magnetic polarons, involving FM clusters
of Eu$^{2+}$ spins just above $T_c$.
It was conjectured \cite{Nyhus} that the transition at higher temperature
arises from the mutual interaction between the magnetic moments and the
conduction electrons which leads to the formation of bound magnetic polarons. 
The bound magnetic polaron corresponds to a composite object 
of localized charge carrier
and its induced alignment in a background of local moments.
On the other hand, Hirsch \cite{Hirsch} proposed a model that 
the magnetism of EuB$_6$ is driven by
the effective mass reduction or the band broadening upon spin polarization. 
The origin of
this effect is the bond-charge Coulomb repulsion which, in a tight binding
model, corresponds to the `off-diagonal' nearest-neighbor 
exchange and pair hopping matrix elements of the Coulomb interaction.
But the parameters for the exchange and pair hopping
considered in his model are unphysical and
the Eu $4f$-states were treated as delocalized electrons
in contrast to the band structure results \cite{Massidda}.
The RKKY interaction among the localized 4$f$ electrons
through itinerant electrons is also considered
to be an origin of ferromagnetism in EuB$_6$\cite{Cooley}.
More recently, a fluctuation-induced hopping model is proposed
as a transport mechanism for the spin
polaron in a paramagnetic background of fluctuating
local moments\cite{Littlewood}.
In this case, temperature dependence of the resistivity 
is obtained to be proportional to $T^{5/2}$ for $k_B T \gtrsim J_{FM}$ and 
to $T$ for $k_{B} T \gg J_{FM}$,
where $J_{FM}$ is coupling between the local moments. 
They claimed that this transport mechanism is valid 
for high-temperature phase of EuB$_6$.
However, no $T^{5/2}$ behavior is observed and moreover
the crossover temperature is too high to be applicable to EuB$_6$. 

The diverse properties of EuB$_6$ are expected to come primarily
from the exchange interactions between the carriers and Eu$^{2+}$
4$f$ local moments.
Indeed the large energy and field dependence
of the spin-flip Raman scattering peak \cite{Nyhus}
reflects the substantial FM exchange interaction
($J_{cf} \sim 0.1~$eV) between the carrier and the Eu$^{2+}$
spins in EuB$_6$. The magnetic polaron
formation is favored by the large ferromagnetic $J_{cf}$ \cite{Snow},
as a trapped charge lowers its energy by polarizing the local moments
of Eu$^{2+}$.

The low-carrier density magnetic systems such as EuB$_6$ share many properties
with the CMR manganites, such as the insulator-metal transition
concomitant with the FM transition,
large negative magnetoresistance, and the
 formation of magnetic clusters near $T_c$. 
Distinctly from manganites, however, Eu-based systems do not manifest
the structural complexities originating from the strong electron-lattice 
coupling associated with Jahn-Teller effect.
Due to the structural simplicity, EuB$_6$ is an ideal system
for investigation of the interplay between magnetic and transport properties. 
An essential aspect for a theoretical
understanding of the properties of EuB$_6$ is the temperature dependences
of the resistivity.
So it would be of great interest to study 
the temperature and field dependence of resistivity of EuB$_6$.
For this purpose, we consider the spin-polaron Hamiltonian 
taking into account the magnetic excitations 
(magnons in the linear spin wave approximation) 
together with lattice excitations (phonons):
\begin{eqnarray}
H &=&  - \sum_{<i,j>,\sigma} t  c_{i \sigma}^{\dag} c_{j \sigma} 
+  \sum_{\vec{q}}  \omega_{\vec{q}}  a_{\vec{q}}^{\dag} a_{\vec{q}}  
+  \sum_{\vec{p}}  \Omega_{\vec{p}}  b_{\vec{p}}^{\dag} b_{\vec{p}}  
\nonumber \\
&-& 
\sum_{i,\vec{q} } \frac{J_{\vec{q}}}{2} 
e^{i {\vec{q}}.{\vec{R_{i}}} } 
(c^{\dag}_{i\uparrow} c_{i\downarrow} a_{-\vec{q}}^{\dag} 
 +c^{\dag}_{i\downarrow} c_{i\uparrow}  a_{\vec{q}} ) 
\nonumber \\
&+&  \sum_{i,\vec{p} } g_{\vec{p}}    
e^{i\vec{p}.\vec{R_{i}} } 
(c^{\dag}_{i\uparrow} c_{i\uparrow} +c^{\dag}_{i\downarrow} c_{i\downarrow}  )
(b_{\vec{p}} + b_{-\vec{p}}^{\dag}),
\end{eqnarray} 
where $c_{i\sigma}$ is the annihilation operator for the conduction electron with spin 
$\sigma$ at site $i$, 
$a_{\vec{q}}$ and $b_{\vec{p}}$ are the annihilation operators 
for a localized spin (magnon) with momentum $\vec{q}$ and for the phonons
with wave vector $\vec{p}$, respectively, and
$\omega_{\vec{q}}$ and $\Omega_{\vec{p}}$ are the magnon
and phonon frequencies, respectively.
$\omega_q \sim J_{FM}S$ where $J_{FM}$ being the direct FM exchange
interaction between Eu 4$f$ spins ($S$).
The last two terms are the interaction between the itinerant electrons
and localized spins and that between
the local density of the carrier ($n_{i}$) and localized charge vibrations
(phonons), respectively. The non-spin flip interaction term 
($S^{z}(n_{i\uparrow} - n_{i\downarrow})$) which just shifts
the on-site energy is not considered, as we are
interested only in the transport properties. Without loss of generality,
the Coulomb interaction can be neglected in the very low carrier density
limit.

In EuB$_6$, it was observed that clusters of Eu$^{2+}$ spins are formed
via the strong ferromagnetic $c$-$f$ exchange interaction \cite{Nyhus}.
Hence one can assume a strong exchange coupling which induces
the locally ferromagnetically ordered spin clusters in the ground state. 
Formation of magnetic polaron
is possible in the low carrier density magnetic system if the degrees
of spin disorder is sufficient to localize the carrier, but not too
high to prevent the local FM alignment \cite{Snow}.
As for the electron-phonon ($e$-ph) interaction,
the study of Boron isotope effect suggests that
the polarons formed in EuB$_6$ are predominantly magnetic,
with a negligible lattice contribution \cite{Snow}.
It is, however, not so certain 
that the magnetic polaron in EuB$_6$ do not
couple to the lattice degrees of freedom of Eu ions. 
Further, thermal conductivity measurement on EuB$_6$ reveals 
that the $e$-ph scattering is strongly favored below $T_c$ 
to describe the $T^2$ variation of the thermal conductivity \cite{Von}.
Therefore, to explore the role of lattice in
the magnetic polaron formation, it is worthwhile to consider the 
$e$-ph interaction in the Hamiltonian.
 
To decouple the electron-magnon interaction term, we employ
the following canonical transformation, 
\begin{eqnarray}
\tilde{H} &=&  e^{R_{1}}  ~H~  e^{-R_{1}}, \nonumber\\
R_{1} &=& 
\sum_{i,\vec{q}}  \frac{J_{\vec{q}}}{\omega_{\vec{q}}} 
e^{i\vec{q}.\vec{R_{i}} } 
(c^{\dag}_{i\uparrow} c_{i\downarrow} +c^{\dag}_{i\downarrow} c_{i\uparrow}  )
(a_{\vec{q}} - a_{-\vec{q}}^{\dag}).
\end{eqnarray}
In the presence of $e$-ph coupling, spread and depth of 
lattice deformation can be studied by using the 
Lang-Firsov (LF) transformation \cite{LF},
\begin{eqnarray}
\tilde{H}_{\rm LF} &=&   e^{R_2} ~\tilde{H}~ e^{-R_2}, \nonumber\\
R_2  &=& -\sum_{j,\vec{p}}  \frac{g_{\vec{p}}}{\Omega_{\vec{p}}} 
e^{i\vec{p}.\vec{R_{j}} } 
n_j (b_{\vec{p}} - b_{-\vec{p}}^{\dag}).
\end{eqnarray}
Note that, for weak to intermediate $e$-ph coupling, the
variational LF transformation \cite{Das} would give more satisfactory results 
than LF transformation. 
In the present work, we consider the LF transformation for simplicity. 

\begin{figure}[t]
\centering
\includegraphics[scale = 0.35,angle=270]{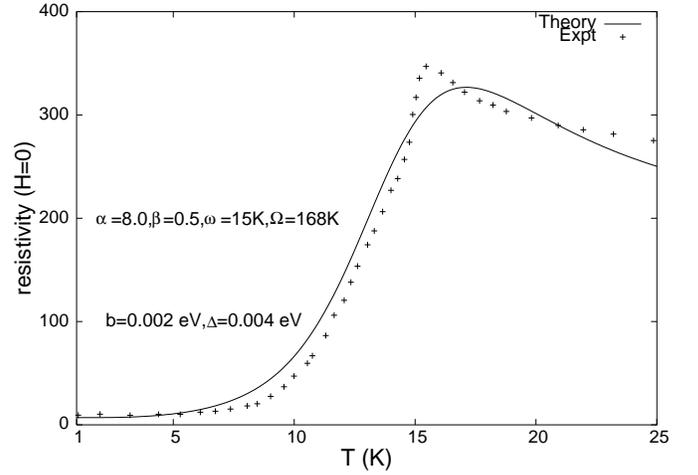}
\caption{ The normalized resistivity (arbitrary unit) for EuB$_6$
in the low-temperature region
with $\omega$=15 K, $\Omega$=168 K, 
$\alpha$=8.0, $\beta$=0.5, $\Delta$=0.004 eV, $b$=0.002 eV. 
`+' symbols represent the experimental data in $\mu \Omega$cm (Ref. \cite{SSullow}).}
\end{figure}

 As a result, the transformed Hamiltonian becomes,
\begin{eqnarray}
\label{Hlf}
\tilde{H}_{\rm LF} &=&  - \sum_{<i,j>,\sigma} t \left[
\cosh(x_i-x_j) X_i^{\dag}X_j c_{i, \sigma}^{\dag} c_{j, \sigma} 
\right.
\nonumber\\
&+& \left.\sinh(x_i-x_j)  X_i^{\dag}X_j c_{i, \sigma}^{\dag} c_{j,-\sigma} 
\right]
+  \sum_{\vec{q}}  \omega_{\vec{q}}  a_{\vec{q}}^{\dag} a_{\vec{q}}  
\nonumber\\
&+& {\cal{O}} (n_{l\uparrow} n_{l\downarrow} )
+ {\cal{O}} (c_{l\uparrow}^{\dag} c_{m\downarrow}^{\dag} 
c_{m\uparrow} c_{l\downarrow})
+   \sum_{\vec{p} } \Omega_{\vec{p}}  b_{\vec{p}}^{\dag} b_{\vec{p}}  
\nonumber \\
&-&  \sum_{i,\vec{p}}  
\frac{g_{\vec{p}}^{2}}{\Omega_{\vec{p}}} n_{i}
-  \sum_{i,j,\vec{p}}^{i \ne j}  \frac{g_{\vec{p}}^{2}}{\Omega_{\vec{p}}} n_{i}n_{j}
e^{i\vec{p}.(\vec{R_i}-\vec{R_j})},\\
{\rm where}~x_i &=& \sum_{\vec{q}}  \frac{J_{\vec{q}}}{\omega_{\vec{q}}} 
e^{i\vec{q}.\vec{R_{i}} } 
(a_{\vec{q}} - a_{-\vec{q}}^{\dag}), \\
X_i &=& \exp{\left[\sum_{\vec{p}}  \frac{g_{\vec{p}}}{\Omega_{\vec{p}}} 
e^{i\vec{p}.\vec{R_{i}} } 
(b_{\vec{p}} - b_{-\vec{p}}^{\dag})\right]}.
\end{eqnarray}
The magnon part remains unchanged after the transformations.
In the transformed Hamiltonian, only the hopping term contains
non-linear functions of spin-wave operators.
As we are interested in the transport properties
in the very low carrier density limit,
we will not consider the renormalized interactions between carriers.
To calculate the electrical conductivity, dynamics of
$\cosh(x_i-x_j)X_i^{\dag} X_j$ and $\sinh(x_i-x_j)X_i^{\dag} X_j$
in the hopping term of Eq. (\ref{Hlf}) should be properly treated. At low
temperature, transport is described by an effective bandwidth 
with a background of magnons and phonons. The effective mass of the
carrier increases as a result of the interaction with the magnons.
The effect of the background is included by calculating the thermal average of 
$\cosh{(x_i-x_j)}X_i^{\dag} X_j$ and $\sinh{(x_i-x_j)}X_i^{\dag} X_j$.
 
\begin{figure}[t]
\centering
\includegraphics[scale = 0.35,angle=270]{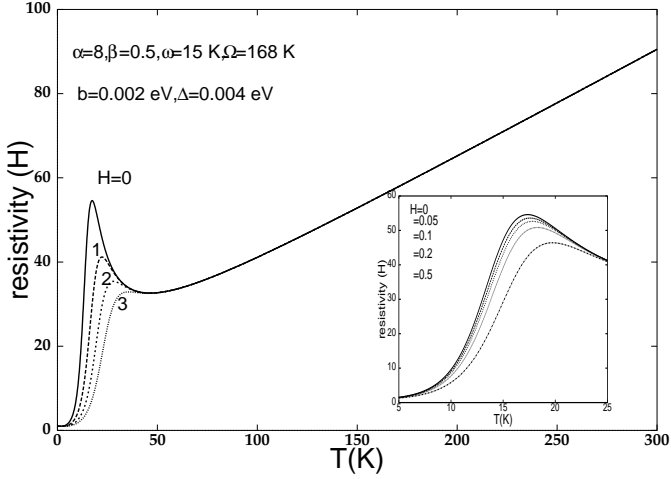}
\caption{
Normalized resistivity vs. temperature $T $ (K) for
different magnetic fields $H $ (in Tesla). Inset figure
shows the low-temperature variation of resistivity.
}
\end{figure}

In the conduction, there are
two independent processes: elastic and inelastic.
 For the elastic conduction process, thermal average of the terms mentioned
above is calculated as in the conventional 
polaron problem \cite{Mahan}, assuming all magnons 
and phonons to be independent:
\begin{eqnarray}
\langle \cosh{(x_i-x_j)}X_i^{\dag} X_j \rangle = 
\exp{\left[- | V_{\vec{q}} |^2 N_{\vec{q}}^{{\rm mag}} \right]} 
\nonumber \\
\exp{\left[- | U_{\vec{p}} |^2 N_{\vec{p}}^{{\rm ph}} \right]}
\exp{\left[ - \frac{1}{2} (| V_{\vec{q}}|^2 + | U_{\vec{p}}|^2) \right]}, 
\end{eqnarray}
and $\langle\sinh{(x_i-x_j)} X_i^{\dag} X_j \rangle=0$.
Here $N_{\vec{q}}^{{\rm mag}}$ and 
$N_{\vec{p}}^{{\rm ph}}$ are the magnon and phonon numbers, respectively,
\begin{eqnarray}
N_{\vec{q}}^{{\rm mag}} &=& \left[ 
\exp[\omega_{\vec{q}}/(k_{B}T)]-1 \right] ^{-1}, \nonumber \\
N_{\vec{p}}^{{\rm ph}} &=& \left[ 
\exp[\Omega_{\vec{p}}/(k_{B}T)]-1 \right] ^{-1}, \\
{\rm and}~~~V_{\vec{q}} &=& \frac{J_{\vec{q}}}{\omega_{\vec{q}}}
e^{i\vec{q}.\vec{R_{i}} } 
(1 - e^{i{\vec{q}}.{\vec{\delta}} }),
\nonumber \\
U_{\vec{p}} &=&  \frac{g_{\vec{p}}}{\Omega_{\vec{p}}} 
e^{i\vec{p}.\vec{R_{i}} } 
(1 - e^{i{\vec{p}}.{\vec{\delta}} }),
\end{eqnarray}
where $\vec{\delta}=\vec{R}_i -\vec{R}_j$ and $\langle ...\rangle$ denotes
the thermal average. The temperature dependence in the
above thermal average comes only through 
$N_{\vec{q}}^{{\rm mag}}$ and $N_{\vec{p}}^{{\rm ph}}$.
If we assume at finite temperature that the metallic conduction
of the elastic process $\sigma_{e}$
is inversely proportional to the effective
mass \cite{Zhang} and the effective mass is inversely
proportional to the effective bandwidth, then
\begin{equation}
\sigma_e(T) \propto \langle t \cosh(x_i-x_j)X_i^{\dag} X_j  \rangle. 
\end{equation}
 
 For the inelastic process, the conduction is provided 
by the incoherent hopping
with emitting and absorbing the magnons and phonons.
The conductivity is given by the Kubo formula \cite{Mahan},
\begin{equation}
\sigma_{\rm in} = n e^2 w \delta^2/ (3k_B T), 
\end{equation}
where $n$ is density of mobile carriers, and $w$ is the transition probability
rate given by the Fermi golden rule,
\begin{equation}
w = (t^2/\hbar \Delta) \exp(-\Delta/k_BT), 
\end{equation}
with the activation energy 
$\Delta = \frac{1}{4} (\sum_{\vec{q}} \omega_{\vec{q}} |V_{\vec{q}}|^{2}  
+ \sum_{\vec{p}} \Omega_{\vec{p}} |U_{\vec{p}}|^{2})$.
For simplicity, if we consider single magnon ($\omega$) and
phonon ($\Omega$) frequency,
the total conductivity is given by the sum of 
two independent processes,
\begin{eqnarray}
\sigma(T)/\sigma(0) &=& \exp{[-\alpha N(\omega) -\beta N(\Omega)]}
\nonumber\\
&+& (b^2/(3 \Delta k_B T))\exp{(-\Delta/k_B T)},
\end{eqnarray}
where $\alpha = \sum_{\vec{q}} |V_{\vec{q}}|^{2}$,
$\beta=\sum_{\vec{p}} |U_{\vec{p}}|^{2}$,
$b = e t \delta \sqrt{n/\hbar \sigma(0)}$,
and $\sigma(0)$ is the zero temperature conductivity.

\begin{figure}[t]
\centering
\includegraphics[scale = 0.35,angle=270]{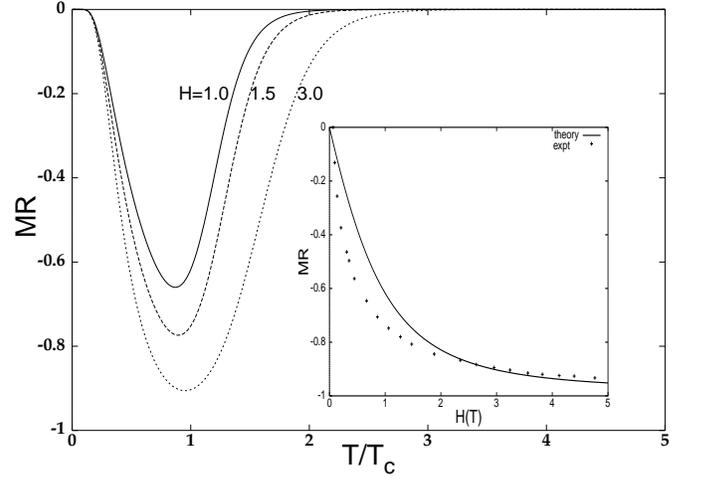}
\caption{ Normalized magnetoresistance (MR) vs.
temperature ($T/T_c$ with $T_c$=15 K) for different values of the 
magnetic field (in Tesla). The same parameters as in Fig.1 are used.
Inset: MR vs. $H$ (in Tesla) at $T$=15~K. `+' symbols represent
the experimental data (Ref. \cite{SSullow}).
}
\end{figure}

For numerical calculation, we consider $\omega \sim T_c$. 
As for $\Omega$, since no inelastic neutron scattering
measurements have been reported on EuB$_6$,
we tentatively use the renormalized frequency
based on the Einstein phonon frequency for LaB$_6$ \cite{Mandrus}.
If we assume $\Omega$=168 K for the localized mode of
Eu$^{2+}$ ion, the model calculation reproduces the experimental features.
The activation energy is a function of $\omega$ and $\Omega$,
and also a function
of $J_{cf}$ and $e$-ph coupling. We observed that nature of the resistivity
is sensitive to the energy parameters $\Delta$ and $b$.
The average distortion around Eu site
is more than an order of magnitude smaller than that in
the perovskites \cite{Booth}, and so
we choose $\beta$ to be much less than $\alpha$.

 Figure 1 provides the calculated resistivity (normalized to $T$=0)
as a function of temperature. The calculated resistivity
describes well the qualitative features of the experimental  
resistivity \cite{Guy,SSullow,Fisk,Rhyee}.
For comparison with the experimental data, we have multiplied
our results by an arbitrary factor and found very satisfactory agreement with 
the experimental results.
In the present model,
the resistivity peak near 15 K signifies the crossover
from a high temperature insulating state
with localized and isolated carriers (magnetic polarons) to a
low temperature conducting state resulting
from the overlap of the magnetic polarons. 
Polaron overlap is marked by the rapid drop in resistivity at the transition. 
Consideration of only the electron-magnon interaction in the present
model can reproduce the resistivity peak around 15 K. 
However, the inclusion of phonons within this model
makes the agreement better. At very low temperature, the contribution of 
magnons to the conductivity is more important than that of phonons.
With increasing temperature, the band conduction dominates with
enhanced effective mass of the carrier so that
the resistivity increases very rapidly.
In contrast, at very high temperature, the role of the lattice effect becomes
significant due to the scattering with the phonons \cite{Von},
where the incoherent hopping dominates.

 Now let us investigate the effect of the external magnetic 
field on the transport. 
One can include the effect of the external magnetic field $H$ for FM magnons 
by replacing the magnon frequency 
$\omega_{\vec{q}}$ by $(\omega_{\vec{q}} +g_{{\rm eff}} \mu_B H)$,
where $g_{{\rm eff}}$ is the effective
$g$-factor for the localized spins of Eu$^{2+}$.
$g_{{\rm eff}}(= g^{*}+ \frac{J_{cf} \chi}{g \mu_B^2 x_c})$ takes into 
account the enhanced effective field arising form the
exchange interaction between the carriers and Eu-4$f$ local moments 
in the presence of the magnetic field.
Here $g^{*}$ is the intrinsic $g$ factor for the Eu-4$f$ spin,
$\chi$ and $g$ are the susceptibility and
the $g$ factor for the conduction electrons, respectively,
and $x_c$ is the concentration of conduction electron.
Taking the temperature independent Pauli paramagnetic susceptibility
for $\chi$ which is given by the density of states $N(E_F)$ of 
conduction electrons \cite{Nef} 
and $x_c=0.01$ per unit cell \cite{Sullow,SullowJAP},
one can estimate the value of $g_{{\rm eff}}$
to be nearly equal to 6. 
Then, with the above field dependent $\omega_{\vec{q}}$,
the experimental results of large magnetoresistance can be achieved.

 The temperature dependence of the electrical resistivity 
with varying the external magnetic field is presented in Fig. 2. 
The field and temperature dependences of the electrical resistivity
reveal that the charge transport is strongly correlated with
the higher transition temperature \cite{Sullow}.
In the presence of magnetic field,
the transition is broadened and shifted to higher temperature.
In the inset shown is the resistivity in the low-temperature region
with varying the external magnetic field ranging from 0.05-0.5 T.
At zero field, the resistance drops sharply just above the transition
temperature. This resistivity peak is suppressed by the magnetic field
in good agreement with the observation \cite{Sullow}.

 In Fig. 3, we plot the magnetoresistance (MR=$(\rho (H)-\rho(0))$$/\rho(0)$)
predicted by the present model as a function of temperature. 
With increasing temperature, a large negative contribution to
MR appears. We obtain the negative MR even at temperature 
higher than $T_c$ which is consistent with the experiment \cite{Guy}.
Inset of Fig. 3 provides the normalized MR at $T$=$T_c$ (15 K)
as a function of the external magnetic field. 
A large negative MR near the FM transition,
as depicted in the inset, is indeed observed in EuB$_6$ \cite{SSullow}.
The MR approaches $-1$ in the high field limit,
reflecting that the zero field  resistivity is completely suppressed,
which is in excellent agreement with experimental features.

In conclusion, we have proposed the model which accounts qualitatively 
well for several anomalous transport properties observed in EuB$_6$.
The strong exchange interaction ($J_{cf}$) and the FM 
magnons in the present model can reproduce the temperature and 
magnetic field dependences of the resistivity in EuB$_6$.
Large negative MR occurs even at temperature higher than $T_c$,
which is consistent with experiments.

Acknowledgements $-$ 
This work was supported by the KOSEF through the eSSC at POSTECH
and in part by the KRF (KRF--2002-070-C00038).

\end{document}